\documentclass[reprint,aps,pre,showpacs,superscriptaddress]{revtex4-1}
\pdfoutput=1
\usepackage{amsmath}
\usepackage{mathptmx}
\usepackage{newtxmath}
\usepackage[T1]{fontenc}
\usepackage[latin9]{inputenc}
\usepackage{bm}
\usepackage{graphicx}
\usepackage{xcolor}

\makeatletter

\newcommand{\expv}[1]{\left\langle #1\right\rangle}

\makeatother

\begin{document}

\title{Thresholding normally distributed data creates complex networks}

\author{George~T.~Cantwell}

\affiliation{Department of Physics, University of Michigan, Ann Arbor, Michigan,
USA}

\author{Yanchen~Liu}

\affiliation{Center for Complex Network Research, Northeastern University, Boston,
Massachusetts, USA}

\author{Benjamin~F.~Maier}

\affiliation{Robert Koch Institute, Nordufer 20, D-13353 Berlin, Germany}

\affiliation{Department of Physics, Humboldt-University of Berlin, Newtonstraße
15, D-12489 Berlin, Germany}

\author{Alice~C.~Schwarze}

\affiliation{Wolfson Centre for Mathematical Biology, Mathematical Institute, University of Oxford, Oxford OX2 6GG, UK}

\author{Carlos~A.~Serv\'an}
\affiliation{Department of Ecology and Evolution, University of Chicago, Chicago,
Illinois, USA}

\author{Jordan~Snyder}

\affiliation{Department of Mathematics, University of California, Davis, California, USA}
\affiliation{Complexity Sciences Center, University of California, Davis, California, USA}

\author{Guillaume~St-Onge}

\affiliation{Département de Physique, de Génie Physique, et d'Optique, Université
Laval, Québec (Québec), Canada, G1V 0A6}

\affiliation{Centre interdisciplinaire de modélisation mathématique de l'Université
Laval, Québec (Québec), Canada, G1V 0A6}
\begin{abstract}
	Network data sets are often constructed by some kind of thresholding
	procedure.  The resulting networks frequently possess properties such
	as heavy-tailed degree distributions, clustering, large connected
	components and short average shortest path lengths. These properties
	are considered typical of complex networks and appear in many contexts,
	prompting consideration of their universality. Here we introduce
	a simple model for correlated relational data and study the network
	ensemble obtained by thresholding it. We find that some, but not all,
	of the properties associated with complex networks can be seen after
	thresholding the correlated data, even though the underlying data are 
	not ``complex''.  In particular, we observe heavy-tailed degree
	distributions, a large numbers of triangles, and short path lengths,
	while we do not observe non-vanishing clustering or community
	structure.
\end{abstract}
\maketitle

\section{Introduction}

Networks are a popular tool for representing and analyzing real-world systems
consisting of entities and their relationships.  They provide a simple yet
intuitive representation for many complex systems. 
In the most basic incarnation, networks are
simple graphs---undirected and unweighted with only one type of node and one
type of edge. The usual picture is that nodes represent some group of objects
(people, neurons, proteins, etc.), and edges represent some kind of interaction
between them (friendship, synapses, binding, etc.)
\cite{Newman18,berg_structure_2004,bullmore_brain_2011,stopczynski_measuring_2014,apeltsin_improving_2011,bullmore_complex_2009,rubinov_complex_2010,sekara_strength_2014}.

In many real-world settings interactions are indicated by real-valued data,
and so creating a simple network requires thresholding, which may take several
forms
\cite{apeltsin_improving_2011,bullmore_complex_2009,langer_problem_2013,rubinov_complex_2010,huang2009network,sekara_strength_2014,stopczynski_measuring_2014,
serrano2009extracting, radicchi2011information, dianati2016unwinding}.  The
most obvious case of thresholding is when a continuous valued data set is explicitly
thresholded by deciding what level of interaction is sufficiently strong to
count as an edge in the network.  A more subtle case is that of experimental
limitation: interactions that exist but are very weak or rare may not be
observed.  Even for binary valued data sets the sampling method may hide an
implicit thresholding mechanism.  For example, one commonly uses a combination
of a yeast-two-hybrid screen and biochemical assays to detect and verify edges
in protein-protein interaction networks. These methods typically do not detect
weak protein-protein interactions \cite{VAYNBERG200622} and are thus equivalent
to applying a threshold on the edge strength in protein-protein interaction
networks.  For another example, consider friendship networks. Most everyday interactions
between people are presumably not strong enough to constitute friendship.  At
what point does a casual acquaintance cross over to the category of friend?
When people list their friends, in a survey for instance, they will implicitly
apply some criteria to filter the friends from the acquaintances.  Nevertheless, an
understanding of the properties one should expect to observe from thresholded
relational data is currently lacking.

In this paper we examine the properties of networks created by thresholding
relational data. To do this we introduce a basic model of the underlying
relational data which is then thresholded to produce edges in the network. The
model is derived from three assumptions: 
\begin{enumerate}
\item all nodes are statistically identical; 
\item any correlations are local; 
\item the underlying relational data are normally distributed. 
\end{enumerate}
All three of these assumptions---which are no doubt violated in real-world
systems---are quite natural for a null model. Assumption 1, that all nodes are
identical, severely constrains what correlation structures are admissible. In
fact, only two free parameters remain in the covariance matrix once this
assumption is made: a local correlation strength between edges that share nodes, and a global
correlation strength between edges that do not share nodes.
Assumption 2 sets the second of these to zero---edges
that do not have a node in common are uncorrelated.  The other free
parameter, the local correlation strength, we call $\rho$.  Our remaining
freedom is to pick a distribution that is consistent with the required
correlation matrix.  The most obvious and simple choice is assumption 3, the
multivariate normal (Gaussian) distribution.  We believe this to be the
simplest non-trivial model for relational data.

The thresholding procedure will also be very simple: any of the relational data
that falls above some threshold, $t$, will be said to constitute an edge in the
network, and any that falls below will not. The threshold value $t$ is
a parameter of the model.

Sophisticated methods to extract networks from weighted data have been
developed, for example in \cite{serrano2009extracting, radicchi2011information,
dianati2016unwinding}.  These more complicated methods lead to different
networks, but in this paper we do not consider the relative merits of more
advanced procedures.  We favor the simplistic approach since it contains only
one parameter and allows us to derive equations for several network properties.
Nevertheless, in Appendix~E we present some similar results for the so-called
disparity filter of \cite{serrano2009extracting}.

Our network ensemble on $n$ nodes is thus defined by two parameters: the
threshold, $t$, and a local correlation coefficient, $\rho$.  Despite the
simplicity of the model---the underlying relational data are normally
distributed---we nonetheless find a number of the behaviors typically observed
in complex networks, such as heavy-tailed degree distributions, short average
path lengths, and large numbers of triangles.  It does not, however, yield
non-vanishing clustering or community structure in the large $n$ limit and so
cannot account for this observation in real-world data sets.  Finally, the
model we study is not constrained to produce positive definite matrices.  As
a result it is not immediately applicable to the study of thresholded
correlation matrices (as for example in \cite{huang2009network}).

This paper has two main parts. In Sec.~\ref{sec:specification} we define and
justify the network model. Then, in Sec.~\ref{sec:properties}, we study the
properties of the network ensemble. We look at the density of edges, triangles
and clustering, the degree distributions, shortest path lengths, and the giant
component.

\section{Model specification}

\label{sec:specification}

\subsection{Thresholding locally correlated data}

A network can be represented by its adjacency matrix, $\bm{A}$, where
$A_{ij}=1$ if node $i$ and $j$ are connected and $A_{ij}=0$ otherwise. We
consider networks created by thresholding underlying relational data, $\bm{X}$,
adding an edge between $i$ and $j$ if 
\begin{equation}
X_{ij}\geq t\label{eq:threshold_condition}
\end{equation}
(see Fig. 1a).
To fully specify the model we need to pick a distribution for $\bm{X}$.
Assuming that all nodes are statistically identical---exchangeable in the
parlance of statistics---constrains our choice of distribution.

\begin{figure*}
	\centering
\includegraphics[width=0.75\linewidth]{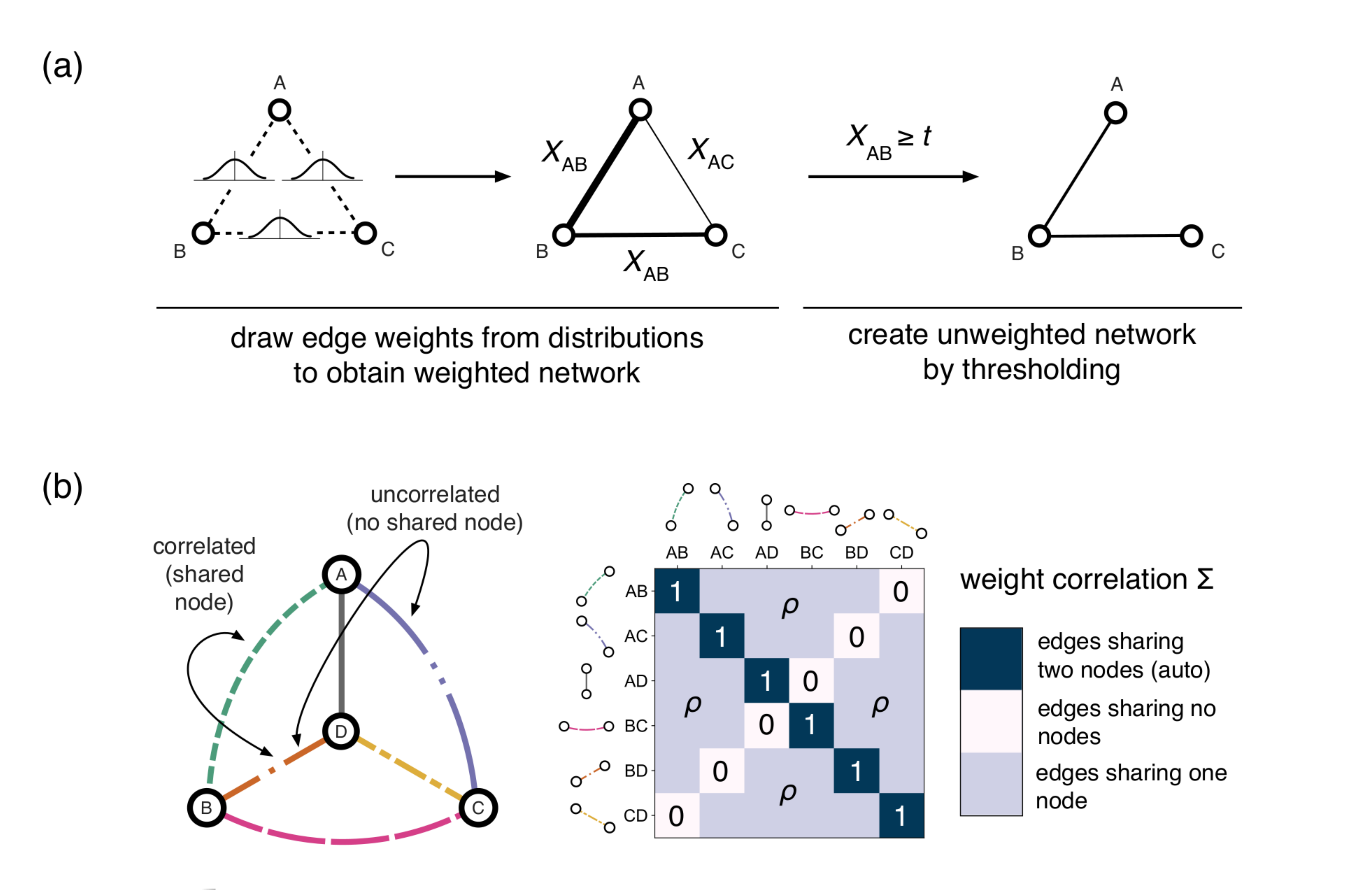}\caption{\label{fig:Thresholding-relational-data}Thresholding
	relational data to obtain networks. Panel (a) shows a general procedure
	to obtain unweighted networks from edge weights. Each edge weight is
	hypothesized to have been drawn from a specific distribution,
	generating an undirected weighted network. An unweighted network is
	then produced by assigning an edge whenever an edge weight $X_{ij}$ is
	greater than a threshold $t$. In panel (b) we show how edge weights are
	correlated in the model of Sec.~\ref{sec:specification} by covariance
	matrix $\bm{\Sigma}$ (Eq.~\eqref{eq:sigma}). Edge weights for edges
	which connect through a node have covariance
	$\mathrm{Cov}[X_{ij}X_{ik}]=\rho$, while edge weights not connected by
	a node have zero covariance.}
\end{figure*}

If nodes are identical then the marginal distribution for $X_{ij}$ must be the
same for all (distinct) pairs $i$ and $j$.  Further, by a linear transform we
can always set $E[X_{ij}]=0$ and \mbox{$\text{Var}[X_{ij}]=1$}. So long as
the appropriate transformation is made to $t$, this shift will have no effect
on the thresholded network. For this reason we will always assume $X_{ij}$ has
mean $0$ and variance $1$. Exchangeability puts further constraints on the
covariance matrix, whose entries can take only three values.  For $i,j,k,l$ all
distinct, these are 
\begin{align}
\text{Var}[X_{ij}] & =\Sigma_{(i,j),(i,j)}=1,\nonumber \\
	\text{Cov}[X_{ij},X_{ik}] & =\Sigma_{(i,j),(i,k)}=\rho, \nonumber \\
\text{Cov}[X_{ij},X_{kl}] & =\Sigma_{(i,j),(k,l)}=\gamma,
	\label{eq:cov_mat}
\end{align}
where $\mathrm{Cov}[X,Y]$ denotes covariance. We will assume that
\mbox{$\gamma=0$} since this quantifies the correlation between two edges that
do not share a node, i.e. two edges that do not ``touch'' (see Fig. 1b).  This leaves us with
two free parameters, $t$ and $\rho$.  The remaining task is to pick
a distribution with the required covariance matrix, $\bm{\Sigma}$.

In principle any distribution could be used, but the obvious choice is
a multivariate normal distribution.  In standard notation a multivariate normal
distribution (MVN) is denoted $\mathcal{N}(\bm{\mu},\bm{\Sigma})$. The
probability density function of an $N$-dimensional MVN is 
\begin{equation}
P\left(\bm{x}\right)=\frac{e^{-\frac{1}{2}(\bm{x}-\bm{\mu})^{T}\bm{\Sigma}^{-1}(\bm{x}-\bm{\mu})}}{\sqrt{(2\pi)^{N}\text{det}(\bm{\Sigma})}}.
\end{equation}

The normal distribution has many points in its favor. Famously it arises in the
central limit theorem, which makes it a plausible model for many random
processes. If the relational data $\bm{X}$ arises due to the aggregation of
many independent processes then the central limit theorem implies $\bm{X}$ will
be multivariate normally distributed.  Further, the normal distribution is the
maximum entropy distribution with the required covariance matrix,
Eq.~\eqref{eq:cov_mat}, and so could be justified as the ``least informative
distribution''---the model that makes the fewest extra assumptions beyond the
correlation structure.  We can also appeal to simple pragmatism: the
multivariate normal distribution is well-studied and has convenient
mathematical properties.

A concise statement of the model is as follows: given the freely chosen parameters
$t\in\mathbb{R},\rho\in{\left[0,\frac{1}{2}\right]}$, and the number of nodes
$n$, draw a random variable $\bm{X}$ with 
\begin{equation}
\bm{X}\sim\mathcal{N}(\bm{0},\bm{\Sigma}),
\end{equation}
where 
\begin{align}
\label{eq:sigma}
\Sigma_{(i,j),(i,j)} & =1,\nonumber \\
\Sigma_{(i,j),(i,k)} & =\rho, \nonumber \\
\Sigma_{(i,j),(k,l)} & =0. 
\end{align}
Then create the network by thresholding $\bm{X}$, 
\begin{equation}
A_{ij}=\begin{cases}
1 & \text{ if }X_{ij}\geq t,\\
0 & \text{ otherwise}.
\end{cases}
\end{equation}
Note, we constrain $0\leq\rho\leq\frac{1}{2}$ so that $\bm{\Sigma}$ is positive
semi-definite \footnote{To see why $\rho>\frac{1}{2}$ is problematic, consider
the marginal distribution for four edges, say $X_{ij},X_{jk},X_{kl},X_{il}$.
A simple calculation shows that the covariance matrix has a negative eigenvalue
for $\rho > \frac{1}{2}$. Similarly, since $\text{Var} \left[ \Sigma_j X_{ij}
\right]$ must be greater than zero, $\rho$ must be greater than $-1/(n-2)$ and
so negative correlations can be vanishingly weak at most.}.

Even if we have good reason to believe that the marginal distributions for
$X_{ij}$ are not normal, the model may still be applicable.  Consider an
arbitrary cumulative distribution function, $F(x)$, and let $\Phi(x)$ denote the
standard normal cumulative distribution function.  If we sample $\bm{X}$ from
a multivariate normal distribution, and then apply the function
$F^{-1}(\Phi(x))$ to each $X_{ij}$ we will have transformed the edge weights to
the arbitrary distribution $F$.  So long as we apply the same transformation to
$t$, however, the resulting network after thresholding will be identical.

The upshot is that our model can be adapted for any marginal distribution, and
no network properties change---the assumption that the edge weights have
normally distributed marginals is of no real consequence.  What \emph{is} important is
the assumption that there is some transformation of the data such that the
\emph{joint} distribution is multivariate normal.  While this assumption is
a limitation, the above procedure is actually one of the standard methods for
creating multivariate distributions with arbitrary marginals.

\subsection{Sampling from the model}

\label{subsec:sampling}

We now describe a simple algorithm to sample from the model.  This algorithm
also provides an intuitive model interpretation.

\noindent \rule{1\columnwidth}{1pt}
Let $Z_{i}$ be $n$ i.i.d.~variables, $\mathcal{N}(0,1)$. Let $Y_{ij}$
be ${n \choose 2}$ i.i.d.~variables, $\mathcal{N}(0,1)$. Then let
\begin{equation}
W_{ij}=\sqrt{1-2\rho}Y_{ij}+\sqrt{\rho}\left(Z_{i}+Z_{j}\right).
	\label{eq:WYZ}
\end{equation}
Note that $W_{ij}$ is normally distributed with mean zero and further
\begin{align}
	\text{Var}\left[W_{ij}\right]&=1, \nonumber \\
	\text{Cov}\left[W_{ij},W_{ik}\right]&=\rho, \nonumber \\
	\text{Cov}\left[W_{ij},W_{kl}\right]&=0. 
\end{align}
Hence, $\bm{W}$ is distributed identically to $\bm{X}$. So, to sample from the
model:
\begin{enumerate}
\item Sample $\bm{z}$, a length n vector of i.i.d.~standard normal variables. 
\item For $i<j$, generate $y\sim\mathcal{N}(0,1)$, and if 
\begin{equation}
y>\frac{t-\sqrt{\rho}\left(z_{i}+z_{j}\right)}{\sqrt{1-2\rho}}
\end{equation}
add edge $(i,j)$ to the network. 
\end{enumerate}
If $\rho=\frac{1}{2}$, generating $y$ is unnecessary and one can simply add
edge $(i,j)$ if $\sqrt{1/2}(z_{i}+z_{j})\geq t$.

\noindent \rule{1\columnwidth}{1pt}

A Python package to generate networks along with scripts for the figures in
this paper is publicly available~\cite{maier_thredgecorr_2018}. 

In order to achieve the required correlations, the algorithm above separates
$X_{ij}$ into node and edge effects. Each node is given a value $Z_{i}$ and
$X_{ij}$ is created by a linear combination of $Z_{i}$ and $Z_{j}$ plus i.i.d.
random noise $Y_{ij}$. We can interpret the $Z$'s as latent variables that
control the propensity for individual nodes to have edges and $\rho$ controls
the relative strength of the noise process. When $\rho=1/2$ edges are
entirely determined by the values of $Z$, while at $\rho=0$ edges are entirely
random and independent. 

Despite this equivalent formulation, our model should not be primarily
understood as a latent variable model since it was not constructed as one.
Rather, the equivalent latent variable model is derived and used for
algorithmic convenience.  In fact, the existence of this latent variable
interpretation is not surprising.  As $n \to \infty$ our model is in
a class of models known as \emph{exchangeable random graphs}
\cite{diaconis2007graph, orbanz2015bayesian}. The Aldous-Hoover theorem implies
that all exchangeable random graphs have an equivalent latent variable model
\cite{diaconis2007graph, orbanz2015bayesian, hoover1979relations}.

\section{Network properties}

\label{sec:properties} We now turn our attention to properties of the networks
created by the model.

\subsection{Edge density}

Edges in the network exist whenever the corresponding weight $X_{ij}$ is
greater than $t$. The marginal distribution for $X_{ij}$ is simply a standard
normal distribution. Thus, 
\begin{equation}
E[A_{ij}]=P[A_{ij}=1]=P[X_{ij}\geq t]=1-\Phi(t),\label{eq:edge_existence}
\end{equation}
where $\Phi(x)$ is the cumulative distribution function for the standard normal
distribution $\mathcal{N}(0,1)$. When $\rho=0$ all edges exist independently
and the model is equivalent to the random graph, $G_{n,p}$, with $p=1-\Phi(t)$.

The mean degree is equally simple to compute. For all $\rho$
\begin{equation}
E[k_{i}]=\sum_{j\neq i}E[A_{ij}]=(n-1)(1-\Phi(t)).\label{eq:Ek}
\end{equation}
If we want to pick $t$ for a desired mean degree $\expv k$, it is
easy to invert this to obtain 
\begin{equation}
t=\Phi^{-1}\left(1-\frac{\expv k}{n-1}\right).\label{eq:t_fit}
\end{equation}

\subsection{Triangles, clustering, and degree variance}

\label{subsec:clustering}

Many complex networks are observed to have large numbers of triangles.  The
clustering coefficient or transitivity is one way to quantify this. We can
quantify the clustering with the probability that a triangle is closed, given
that two of its edges already exist, 
\begin{equation}
C=P[A_{ik}=1|A_{ij},A_{jk}=1]=\frac{P[A_{ik},A_{ij},A_{jk}=1]}{P[A_{ij},A_{jk}=1]}.
\end{equation}
The numerator of this equation corresponds to the density of triangles while
the denominator corresponds to the density of two-stars (which also determines
the variance of the degree distribution). Note that for simplicity we shorten
the logical connective ``and'' (or ``$\wedge$'') using commas, e.g.
\mbox{$P[A_{ij}=1\ \wedge\ A_{jk}=1]\equiv P[A_{ij},A_{jk}=1]$}.

\begin{figure*}
	\centering
\includegraphics[width=1\linewidth]{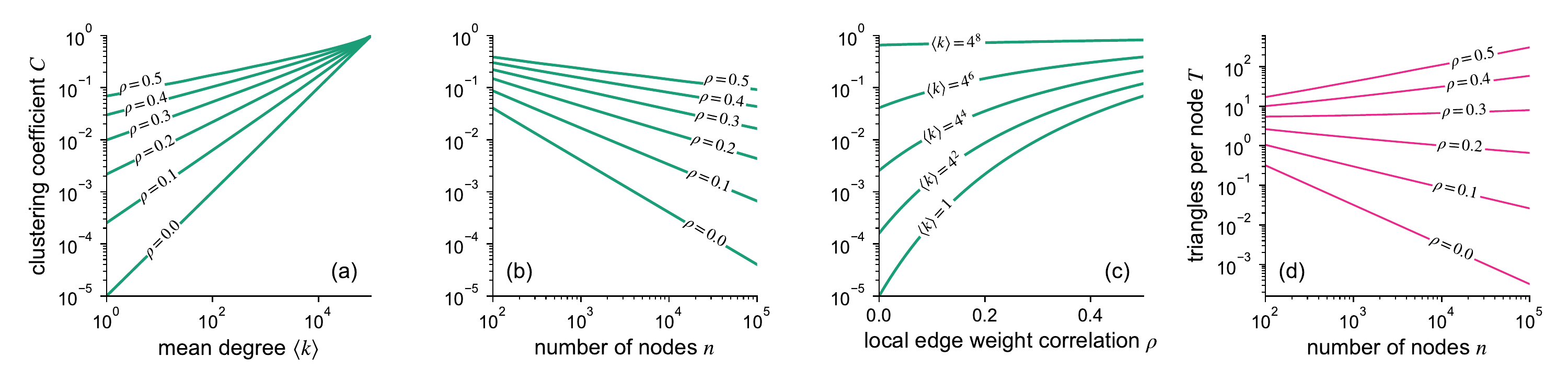}\caption{\label{fig:Clustering--and}Clustering $C$ and triangles per node $T$
as computed in Sec.~\ref{subsec:clustering}. Clustering decreases
with increasing number of nodes, however the number of triangles per node increases
with growing number of nodes $n$ for large values of $\rho$. Clustering
increases both with increasing mean degree $\expv k$ and local edge
weight correlation $\rho$. In panel (a) and (c) we chose $n=100\,000$
and in panel (b) and (d), we fixed $\expv k=4$.}
\end{figure*}

The marginal distributions of a MVN are themselves MVN, and are found by
simply dropping the unwanted rows and columns in the correlation matrix
$\bm{\Sigma}$. Thus, $(X_{ij},X_{ik})^{T}$ will be bivariate normally
distributed and $(X_{ij},X_{ik},X_{jk})^{T}$ will be trivariate normally
distributed, both with correlation coefficient $\rho$. Introducing the Hermite
polynomials $H_{N}(x)$ as defined in Appendix~\ref{sec:hermite_and_two_stars},
one finds that 
\begin{equation}
\label{eq:P2x_hermite}
P[X_{ij},X_{ik}\geq t]=\sum_{N=0}^{\infty}\frac{\rho^{N}}{N!}\left[\phi(t)H_{N-1}\left(t\right)\right]^{2} 
\end{equation}
for the density of two-stars and 
\begin{align}
	&P[X_{ij},X_{ik},X_{jk}\geq t]= \nonumber \\
	&\sum_{N=0}^{\infty}\:\,\sum_{i=0}^{N}\sum_{j=0}^{N-i}\frac{\rho^{N}\phi\left(t\right)^{3}}{i!\:j!\:(N-i-j)!}H_{N-1-i}\left(t\right)H_{N-1-j}\left(t\right)H_{i+j-1}\left(t\right)\label{eq:P3x_hermite}
\end{align}
for triangles. Both sums converge for $\rho\leq0.5$, and we can estimate them
accurately with a finite number of terms~\cite{harris_use_1980}. Noting that
there are ${n-1 \choose 2}$ potential triangles for each node, the expected
number of triangles per node is simply ${n-1 \choose 2}$ times their density 
\begin{equation}
T={n-1 \choose 2}P[X_{ij},X_{ik},X_{jk}\geq t].
\end{equation}
Plots of these functions are shown in Fig. \ref{fig:Clustering--and}.  We find
that $T$ is much larger in these networks than in the random graph
$G_{n,p}$---larger by multiple orders of magnitude.  In fact, while $T$ goes to
zero in the large $n$ limit for the random graph, in this model we find that
$T$ increases with $n$ for large values of $\rho$.  On the other hand, the
clustering coefficient $C$ decreases with growing number of nodes for all
parameter values. This leads to a slightly paradoxical result for large $\rho$:
in the limit $n\rightarrow\infty$ the expected number of triangles at each node
goes to infinity, and the clustering coefficient still goes to zero! The reason
for this is that the number of two-stars diverges faster than the number of
triangles.

\begin{figure}
	\centering
\includegraphics[width=3.1in]{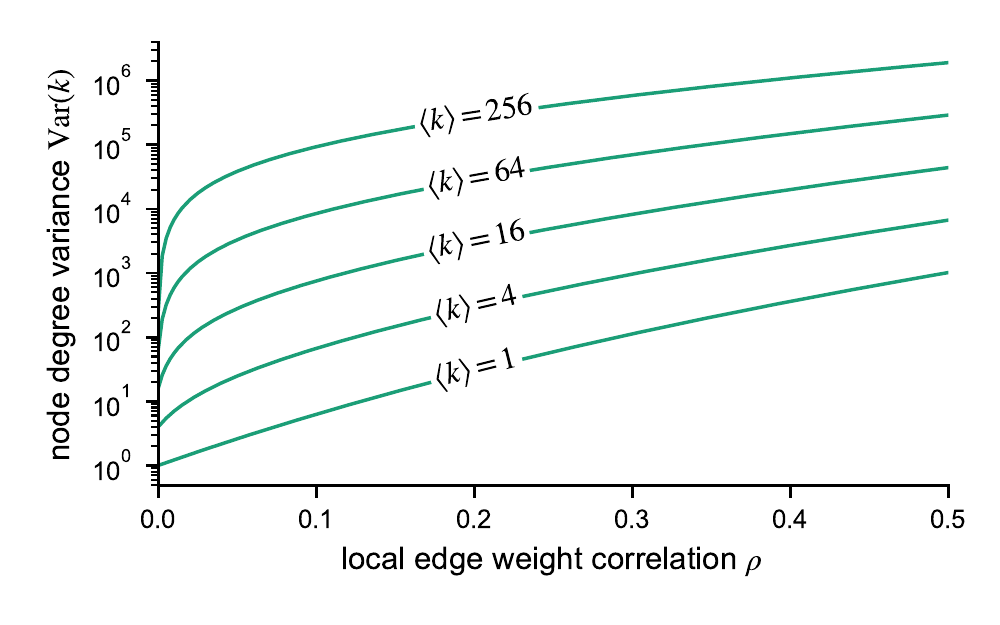}\caption{\label{fig:Vark}The variance of degree, Eq.~\eqref{eq:Vark}, increases
with $\rho$, the local edge weight correlation. With increasing mean
degree $\expv k$, even small correlations $\rho$ produce networks
	of significantly broader degree distribution than the random graph $G_{n,p}$.}
\end{figure}

Equation~\eqref{eq:P2x_hermite} can also be used to compute the variance of the
degree distribution. To see this note that a node of degree $k$ has ${k \choose
2}$ two-stars. Further, noting that there are ${n-1 \choose 2}$ potential
two-stars (the same number of potential triangles) we find
\begin{equation}
\frac{1}{2}\left( \langle k^2 \rangle -\expv{k}\right)={n-1 \choose 2}P[X_{ij},X_{ik}\geq t].
\end{equation}
Combining this with Eq.~\eqref{eq:Ek} the variance of the node degree
$k$ can be written
\begin{align}
	\text{Var}[k]&=  (n-1)  \Phi(t)\left[1-\Phi(t)\right] \nonumber \\
	& + (n-1)(n-2)\sum_{N=1}^{\infty}\frac{\rho^{N}}{N!}\left[\phi(t)H_{N-1}\left(t\right)\right]^{2} .\label{eq:Vark}
\end{align}
The first term is simply the variance of a binomial distribution.  For $\rho=0$
the second term vanishes and we recover the correct result for the random graph
$G_{n,p}$. For $\rho>0$ the sum is positive and monotonically increases with
$\rho$ as illustrated in Fig.~\ref{fig:Vark}.

\subsection{Degree distribution}

\label{subsec:degree-distribution}

In the previous two subsections we gave expressions for the mean and variance
of the degrees. Here we give expressions for the full distribution of degrees. 

The degree distribution $p_{k}$ is the probability that a node has $k$ edges.
For this model the degree distribution can be written 
\begin{equation}
\label{eq:degree_dist_integral}
p_{k}  ={n-1 \choose k}\sqrt{\frac{1-\rho}{2\pi\rho}}\int_{-\infty}^{\infty} e^{ f_{k} \left( y \right) }dy
\end{equation}
where 
\begin{align}
	f_{k}(y) = k &\ln \left[1- \Phi(y)\right] +(n-k-1)\ln\left[\Phi(y)\right] \nonumber \\
	&-\frac{1}{2}\left(\frac{t-\sqrt{1-\rho}y}{\sqrt{\rho}}\right)^{2}.
\end{align}
This result is derived in Appendix~\ref{sec:derivation-of-degree-distribution}.

The integral in Eq.~\eqref{eq:degree_dist_integral} can be computed numerically
to high precision using Gauss-Hermite quadrature, centered at the maximum of
$f_{k}(y)$.  Increasing the order of Gauss-Hermite quadrature (i.e.
incorporating more points) increases the accuracy. The full details are in
Appendix~\ref{sec:derivation-of-degree-distribution}.

We can also approximate the integral using Laplace's
method~\cite{miller2006applied}, an asymptotic approximation for integrals of
this form (equivalent to a first order Gauss-Hermite quadrature).  The idea of
the method is to replace the function $f_{k}(y)$ by a second order Taylor
series around its maximum. For large $n$, the last (quadratic) term in $f_{k}$
will be negligible and for $0<k<n-1$, the maximum will be at 
\begin{equation}
y_{0,k}=\Phi^{-1}\left(1-\frac{k}{n-1}\right).
\end{equation}
Combining this with Stirling's approximation for the binomial coefficient, we
find 
\begin{equation}
p_{k}\sim\frac{1}{n-1}\sqrt{\frac{1-\rho}{\rho}}\exp\left[-\left(\frac{1-2\rho}{2\rho}\right)y_{0,k}^{2}+\left(\frac{t\sqrt{1-\rho}}{\rho}\right)y_{0,k}-\frac{t^{2}}{2\rho}\right].\label{eq:asymptotic_pk}
\end{equation}
Together with the closed form approximation for $\Phi^{-1}$, given in
Appendix~\ref{sec:approximation_of_inverse}, Eq.~\eqref{eq:asymptotic_pk}
provides a closed form approximation for the degree distribution. 

\begin{figure*}
	\centering
\includegraphics[width=1\linewidth]{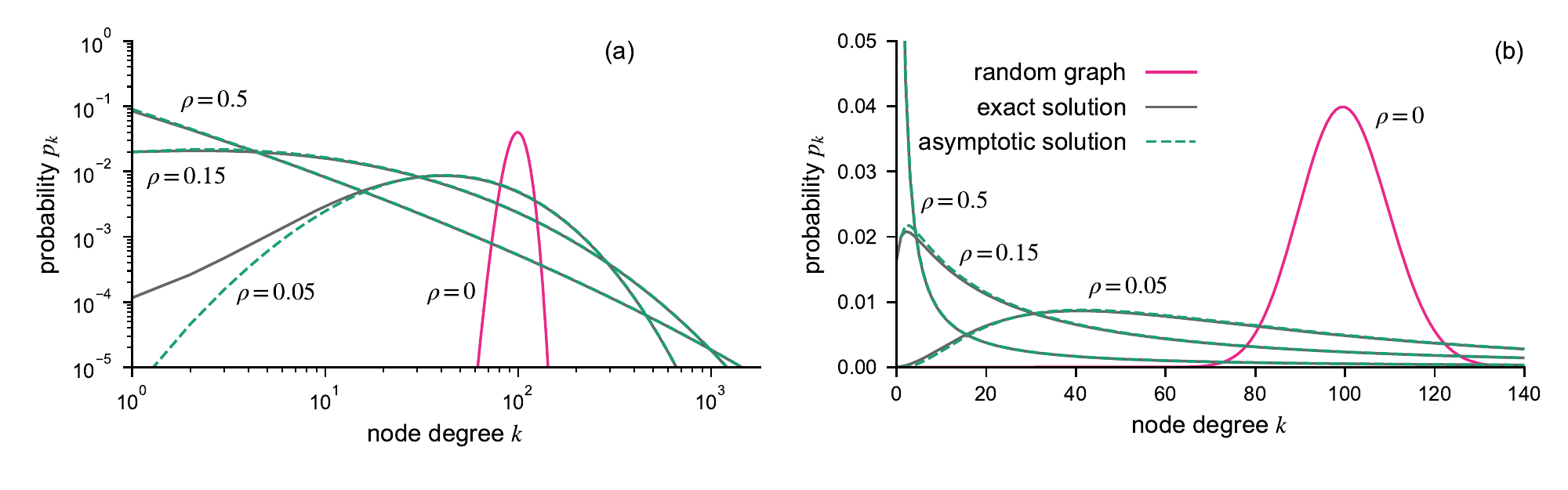}\caption{\label{fig:example_distributions}We show degree distributions computed
using Eq.~\eqref{eq:degree_dist_integral} for $n=100\,000$ and $\expv k=100$
for increasing local edge weight correlation $\rho$  in log-log (a)
and linear scales (b). We also compare them to the asymptotic approximation
Eq.~\eqref{eq:asymptotic_pk}. Note that large values of $\rho$ produce
broad degree distributions which could be easily mistaken for log-normal
or power-law distributions.}
\end{figure*}

Figure \ref{fig:example_distributions} shows some example degree distributions,
computed to high precision using Eq.~(\ref{eq:degree_dist_integral}) along with
the asymptotic approximation, Eq.~\eqref{eq:asymptotic_pk}, where we chose
$n=100\,000$ and $\expv k=100$.

To illustrate how these degree distributions compare to the degree
distributions of real networks, we chose three data sets from different
domains, and fit the model. The first data set is a network of friendships
between students at a U.S. high school \mbox{($n=2587$)}~\cite{AddHealth}, the
second data set is a co-authorship network of researchers
($n=16726$)~\cite{newman_structure_2001}, and the third network describes
interactions between  proteins ($n=6327$)~\cite{joshi-tope_reactome:_2005}.

Given a number of nodes $n$ the model studied in this paper has two free
parameters, $t$ and $\rho$. A simple procedure to fit the model to the data
is to choose $t$ and $\rho$ so that the mean and variance of
the model's degree distribution match the observed values. We use
Eq.~\eqref{eq:t_fit} to fix $t$ and subsequently Newton's method to solve
Eq.~\eqref{eq:Vark} for $\rho$.

\begin{figure*}
\centering \includegraphics[width=1\linewidth]{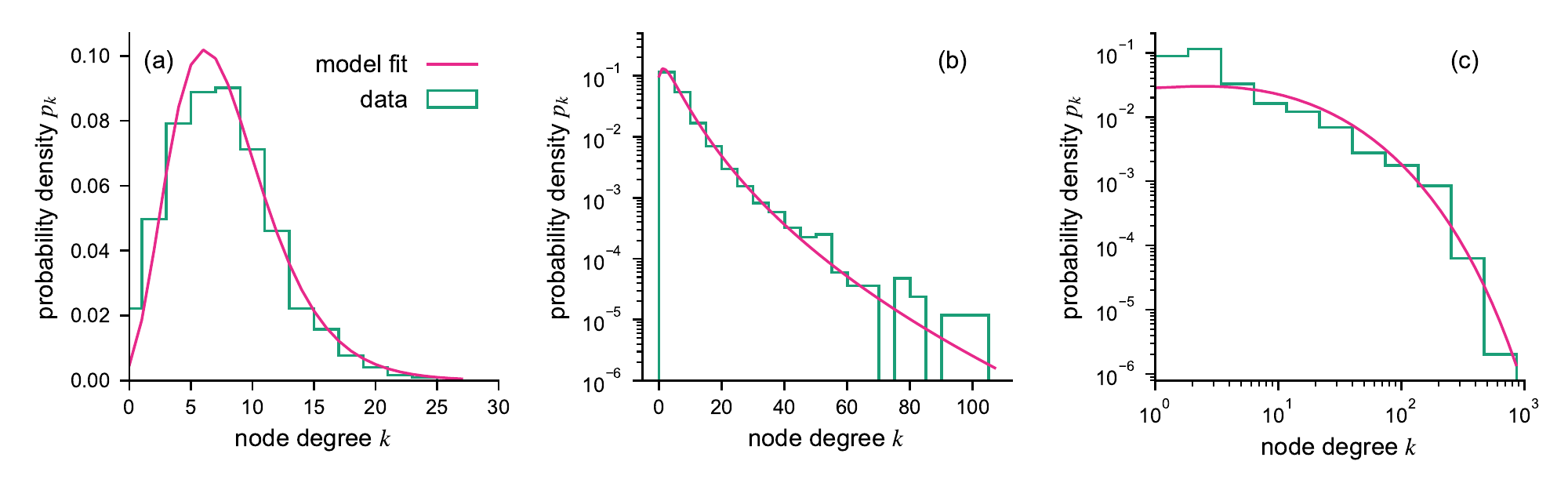}
\caption{Degree histograms for the three real-world networks introduced in Sec.~\ref{subsec:degree-distribution} along with fitted distributions
from the thresholded normal model. We show (a) a high school friendship
network, (b) a co-authorship network between scientists, and (c) a
protein--protein interaction network.}
\label{fig:real_comparisons} 
\end{figure*}

The results of this exercise are shown in Fig.~\ref{fig:real_comparisons}.  The
networks were chosen for their different degree distributions---note the
different scales on the axes: linear, log-linear, and log-log---and
the threshold model can qualitatively ape these distributions.  
Nevertheless, the similarity of the degree distribution should not be over-emphasized.  
As discussed, this model has vanishing clustering so cannot account for this
observation of real-world networks.  Comparisons for clustering are shown in Appendix~D.

While the degrees in the thresholded networks, $k_i = \sum_j A_{ij}$, in
general follow a complicated distribution, the underlying degrees $d_i = \sum_j
X_{ij}$ are always normally distributed.  When $\rho=0$, $d_i$ is Gaussian and
$k_i$ is binomial, or Poisson in the sparse limit.  When $\rho>0$, $d_i$ is
still Gaussian, but $k_i$ now follows a heavy-tailed distribution.  Thus, the
heavy-tailed distribution observed in the model is due to the combination of
correlation and thresholding.  Without positive correlation we observe Poisson
distributions; without thresholding we observe Gaussian distributions.

\subsection{Giant component}

A well studied problem in the theory of random graphs is the formation of
a large connected (giant) component.  At very low densities only a handful of
nodes can be reached from any other node but at some critical point
a macroscopic number of nodes will be connected.  For the random graph this
transition occurs at a mean degree of $\left\langle k\right\rangle =1$
\cite{Newman18, ER60, bollobas_2001}. 

\begin{figure*}
\centering \includegraphics[width=0.7\linewidth]{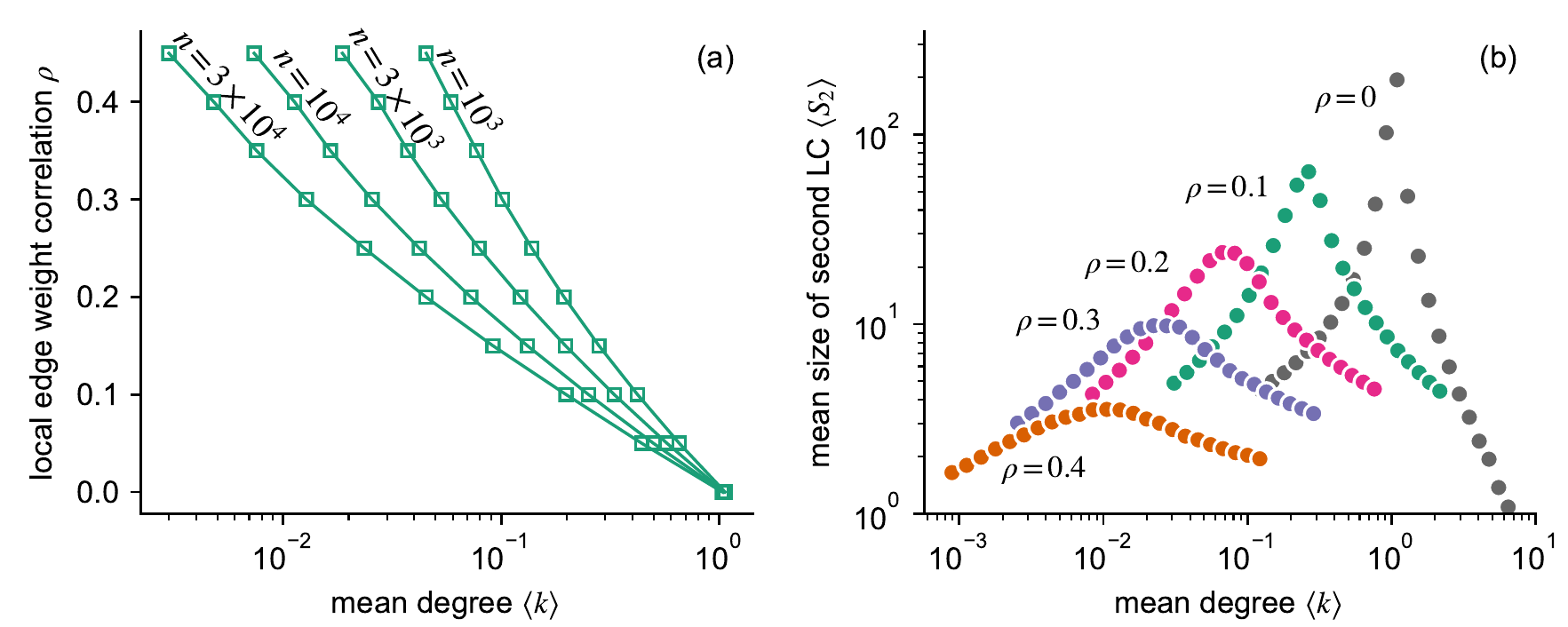}
\caption{
	Simulations with $1\,000\leq n\leq30\,000$, mean degree $10^{-3} \leq \expv k \leq 10$, $0 \leq \rho \leq 0.45$. 1000 samples were taken for each of the parameter combinations. 
	Panel (a) shows the points of transitions for increasing number of
nodes $n$. 
To the left of the line the network does not possess a giant component, 
	while to the right it does.
	The transition point was computed using
the mean size of the second largest component as a susceptibility
parameter. Panel (b) shows an example of the susceptibility parameter
for $n=10\,000$.}
\label{fig:percolation-1} 
\end{figure*}

To explore the effects of $\rho>0$ we sampled from the model as described in
Sec.~\ref{subsec:sampling} and measured the size of the second largest
component as a susceptibility parameter for the phase transition.  The maximum
of this susceptibility parameter is used to find the transition lines in
Fig.~\ref{fig:percolation-1}a.

We find that as $\rho$ or $n$ increases, the transition occurs at lower values
of the mean degree. This result is in line with the configuration model for
which the transition point decreases with increasing variance in the degree
distribution. For $\rho=0$ we recover the standard result for the random graph.

For the other limit case, $\rho=1/2$, recall that all edge weights can be
considered to arise from node ``propensities'', $Z_{i}$, with
$X_{ij}=\sqrt{1/2}(Z_{i}+Z_{j})$. This implies that all nodes that are
connected to any other nodes must also be connected to the node with maximum
propensity $Z_{\max}$. The size of the largest component is then given by this
node's degree plus $1$, $k_{\max}+1$.  The second largest component is then
always of size 1. We therefore omit $\rho=1/2$ in the numerical analysis. 

\subsection{Shortest path lengths}

\begin{figure*}
\centering \includegraphics[width=0.7\linewidth]{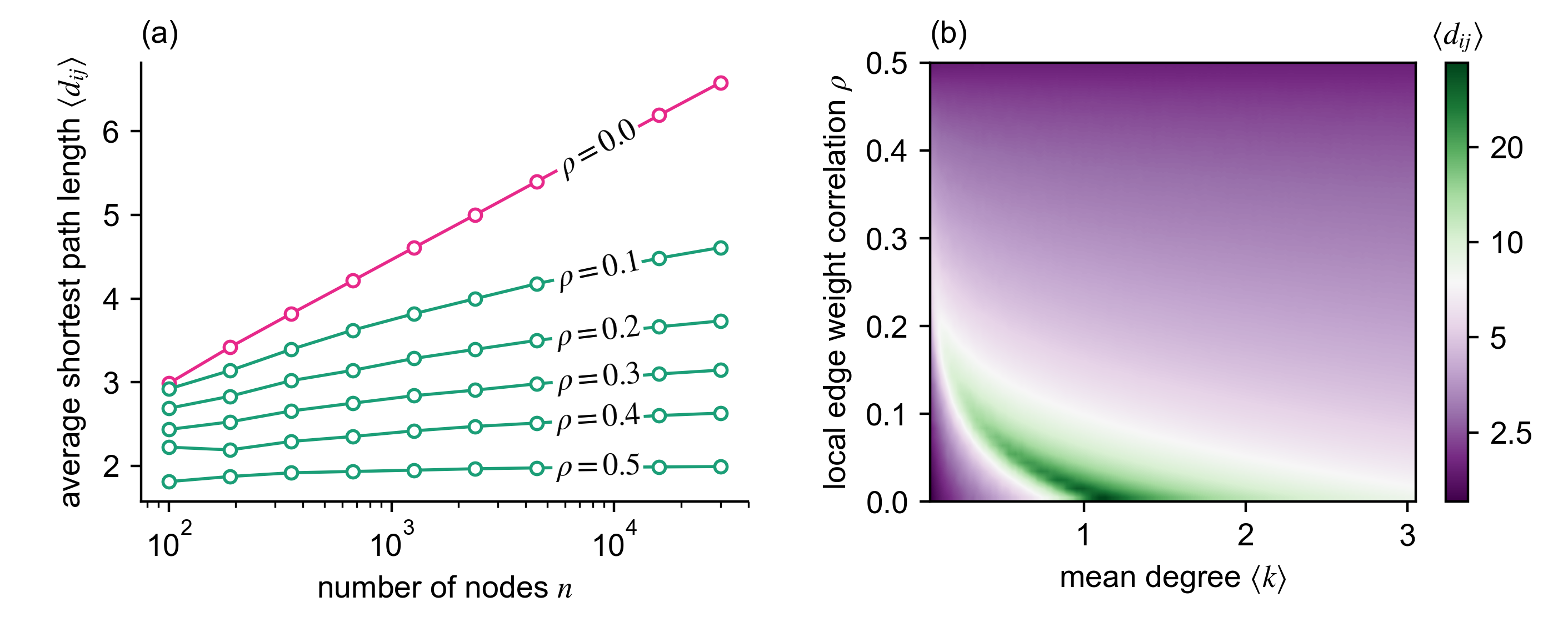}
\caption{Panel (a) shows the scaling of the average shortest path in the largest connected component with the number of nodes, $n$. We fix the mean degree $\expv k=5$ and each point is averaged over 200 samples. For $\rho=0$ we recover the result for the random graph $G_{n,p}$ where
$\left\langle d_{ij}\right\rangle \propto\log n$. 
For non-zero correlation, the average shortest path length increases
slower than logarithmically.
In panel (b) we show the average shortest path length for different
mean degrees and values of $\rho$ for networks with $n=10\,000$,
again sampled 200 times for each parameter combination. 
	}
\label{fig:shortest_paths} 
\end{figure*}

Another phenomenon well established in the complex networks literature is that
randomly chosen nodes often have surprisingly short paths between them. This is
often referred to as the ``six degrees of separation'' or ``small-world''
phenomenon \cite{watts_collective_1998,Newman18}.  By a common definition,
network models are considered to demonstrate this property if the average
shortest path length $\expv{d_{ij}}$ between nodes grows logarithmically (or
slower) as the number of nodes increases \cite{Newman18}. 

Using the method described in Sec.~\ref{subsec:sampling}, we sampled from the
threshold model to verify that it displays this property.  We looked at
networks with between $100$ and $30\,000$ nodes, with mean degree $\expv k=5$,
and investigated the influence of increasing edge weight correlation $\rho$.
After sampling a network from the model we computed the average shortest path
length $\expv{d_{ij}}$ on the largest (giant) component.  For each parameter
combination we computed the mean by averaging 200 sampled networks.

The results are shown in Fig.~\ref{fig:shortest_paths}a. Since it is well known
that the random graph $G_{n,p}$ has short shortest paths \cite{FR07} it is
unsurprising that the threshold model does also (recall, for $\rho=0$ they are
equivalent, and we see the standard $\left\langle d_{ij}\right\rangle
\propto\log n$ scaling behavior). For $\rho>0$ we see that average shortest
path lengths grow significantly slower than logarithmically, a behavior
sometimes referred to as ``ultra small-world'' and often related to networks
with power-law degree distribution
\cite{cohen_scale-free_2003,bornholdt_structural_2004}.  In our model, the
effect appears despite the fact that the degree distribution does not follow
a power-law.

As discussed, when $\rho=1/2$ all edge weights can be considered to arise from
node propensities $Z_{i}$, such that $X_{ij}=\sqrt{1/2}(Z_{i}+Z_{j})$.  All
nodes are then either disconnected or part of the giant component, and the node
with maximum propensity $Z_{\max}$ is connected to all nodes in the giant
component.  Hence, all nodes in the giant component are either directly
connected or can reach each other in two steps through the maximum-degree node.
So, when $\rho=1/2$ the average shortest path length must be
$1\leq\expv{d_{ij}}<2$.

\section{Discussion}
In this paper we studied the effects of thresholding relational data.  We
started with a simple model of multivariate normally distributed data, with
only one free parameter, $\rho$, controlling local correlations.  We then
demonstrated that thresholding this normally distributed correlated relational
data reproduces many of the properties commonly associated with complex
networks. In particular, we find that the combined effects of correlation and
thresholding leads to heavy-tailed degree distributions, relatively large
numbers of triangles, and short average path lengths.

The underlying relational data $\bm{X}$ in the model we introduce would not
usually be considered complex.  It is generated from a highly symmetric
multivariate normal distribution with only one free parameter.  Since every
pair of nodes has some level of interaction, the graphical interpretation for
$\bm{X}$ would be a weighted complete graph, with all edge weights (and linear
combinations thereof) normally distributed.  For example, the ``degrees'', $d_i
= \sum_j X_{i j}$, are normally distributed.  And yet, after thresholding the
networks show several properties commonly associated with complex networks.

One way to think about these results is in the context of the central limit
theorem.  Whenever interaction strengths are the aggregate result of a large
number of processes then we expect $\bm{X}$ to be normally distributed.
Constructing a simple graph from these data can lead to complex networks.  This
provides one simple explanation for the ubiquity of complex networks---they can
arise as a consequence of the central limit theorem.

Of course, for most scientific questions of interest the exact details of the
mechanisms and structure are what matter.  In a social network, for example,
answering the question "who influences whom, and why?" is far from trivial and
the fact that the network has certain commonly observed properties is usually
incidental.

In summary, straightforward assumptions lead to several of the properties
associated with complex networks.  If a network arises by a simple thresholding
procedure then finding that it is ``complex'' need be no more surprising than
finding a bell-shaped curve in a regular data set.

\begin{acknowledgments}
	The authors thank the students, staff, and faculty of the Santa Fe
	Institute Complex Systems Summer School for bringing them together and
	providing a stimulating research environment. All authors made
	significant contributions to the project. G.T.C~and B.F.M.~wrote the
	final manuscript.

	B.F.M.~was supported as an Add-On Fellow for Interdisciplinary Life Science by
	the Joachim Herz Stiftung.  A.C.S.~was supported by the Clarendon Fund,
	e-Therapeutics plc, and the Engineering and Physical Sciences Research
	Council under grant number EP/L016044/1.  J.S.~was supported by the
	U.S.~Army Research Office under Multidisciplinary University Research
	Initiative Award No. W911NF-13-1-0340.  G.S.~was supported by the
	Natural Sciences and Engineering Research Council of Canada and the
	Sentinel North program.

\end{acknowledgments}

\expandafter\ifx\csname url\endcsname\relax \global\long\def\url#1{\texttt{#1}}
\fi \expandafter\ifx\csname urlprefix\endcsname\relax\global\long\def\urlprefix{URL }
\fi\bibliographystyle{ieeetr}
\bibliography{normal_networks}

\appendix

\section{Multivariate normal integrals and Hermite polynomials}

\label{sec:hermite_and_two_stars}The probability of a two-star existing
with nodes $i,j$ and $k$ as constituents is given by 

\begin{equation}
P[X_{ij},X_{ik}\geq t]=\frac{1}{2\pi\sqrt{1-\rho^{2}}}\int_{t}^{\infty}\int_{t}^{\infty}e^{-\frac{1}{2}\left(\frac{x^{2}-2\rho xy+y^{2}}{1-\rho^{2}}\right)}dxdy.\label{eq:P2x-1}
\end{equation}
Direct computation of the integral is not straightforward but
we can compute it quickly using the Hermite polynomials~\cite{harris_use_1980}.
A quick outline of this method: for $n\geq0$, define the Hermite polynomials as 
\begin{equation}
H_{n}(x)=(-1)^{n}e^{\frac{x^{2}}{2}}\frac{d^{n}}{dx^{n}}e^{-\frac{x^{2}}{2}}.
\end{equation}
As the name suggests, the Hermite polynomials are in fact polynomials,
for example $H_{0}(x)=1$, $H_{1}(x)=x$, \mbox{$H_{2}(x)=x^{2}-1$}, and
so on. For notational convenience also define 
\begin{equation}
H_{-1}(x)=\frac{1-\Phi\left(x\right)}{\phi\left(x\right)}.
\end{equation}
Using the Hermite polynomials, we can expand Eq.~\eqref{eq:P2x-1}
as an infinite sum and integrate term by term. The final result is
given by Eq.~\eqref{eq:P2x_hermite}. The same trick is used for
the 3-dimensional integral to give Eq.~\eqref{eq:P3x_hermite}.

\section{Degree distribution}

\label{sec:derivation-of-degree-distribution}Since, by assumption,
all nodes in this model are equivalent, we will simply consider the
one-node marginal to compute the degree distribution. Let $\bm{U}$
be all the terms in $\bm{X}$ that are associated with node $0$,
i.e. $U_{j}=X_{0j}$. Then, $\bm{U}$ is multivariate normally distributed,
$\mathcal{N}(\bm{0},\bm{\Sigma}^{(0)})$, where $\bm{\Sigma}^{(0)}$
has ones along the diagonal and $\rho$ everywhere else 
\[
\Sigma_{jk}^{(0)}=\Sigma_{(0,j),(0,k)}=\begin{cases}
1 & \mathrm{for}\ j=k,\\
\rho & \mathrm{otherwise}.
\end{cases}
\]
The focal node will have degree $k$ when exactly $k$ terms in $\bm{U}$
are larger than the threshold $t$. There are ${n-1 \choose k}$ different
ways this can happen and each is equally likely. So, to compute $p_{k}$
we can compute the probability that the first $k$ terms in $\bm{U}$
are larger than $t$ and all others are smaller, and then multiply
by ${n-1 \choose k}$ to obtain 
\begin{equation}
p_{k}={n-1 \choose k}P\left[U_{1},\dots,U_{k}\geq t;U_{k+1},\dots U_{n-1}<t\right].
\end{equation}
To solve this integral we use a standard trick \cite{tong_multivariate_1990}.
First, we note that if $Z_{0},Z_{1},\dots,Z_{n-1}$ are i.i.d. $\mathcal{N}(0,1)$
then 
\begin{equation}
((\sqrt{1-\rho}Z_{1}+\sqrt{\rho}Z_{0}),\dots,(\sqrt{1-\rho}Z_{n-1}+\sqrt{\rho}Z_{0}))^{T}
\end{equation}
will be distributed identically to $\bm{U}$. Further, once we know
the value of $Z_{0}$ then all the terms are independent, and the
probability that any one of them is greater than $t$ is the probability
that $Z_{1}\geq\frac{t-\sqrt{\rho}z}{\sqrt{1-\rho}}$. Given $Z_{0}=z$,
the probability that exactly $k$ values will greater than $t$ and
the rest less than $t$ is 
\begin{equation}
{n-1 \choose k}\left[1-\Phi\left(\frac{t-\sqrt{\rho}z}{\sqrt{1-\rho}}\right)\right]^{k}\Phi\left(\frac{t-\sqrt{\rho}z}{\sqrt{1-\rho}}\right)^{n-1-k}.
\end{equation}
Averaging this quantity over $z$ then provides us with the correct
expression, 
\begin{align}
p_{k} & ={n-1 \choose k}\underbrace{\int\limits _{-\infty}^{+\infty}\left[1-\Phi\left(\frac{t-\sqrt{\rho}z}{\sqrt{1-\rho}}\right)\right]^{k}\Phi\left(\frac{t-\sqrt{\rho}z}{\sqrt{1-\rho}}\right)^{n-1-k}\phi(z)dz}_{=I_{n,k}}
\end{align}
where $I_{n,k}$ is the integral. A change of variables allows us
to write 
\begin{equation}
I_{n,k}=\sqrt{\frac{1-\rho}{2\pi\rho}}\int_{-\infty}^{\infty}e^{f_k \left( y \right)}dy
\end{equation}
where 
\begin{align}
	f_k(y)=k\ln\left[1-\Phi(y)\right]+(n&-k-1)\ln\left[\Phi(y)\right] \nonumber \\
	-\frac{1}{2}\left(\frac{t-\sqrt{1-\rho}y}{\sqrt{\rho}}\right)^{2}. \nonumber \\
\end{align}

A standard approach to approximate such an integral is to use Laplace's
method. In this approach one expands $f$ about its maximum and then
neglects higher order terms, $f(y)\approx f(y_{0})-\frac{|f''(y_{0})|}{2}(y-y_{0})^{2}$.
Having done this, the integral reduces to a standard Gaussian integral.
While this approach is asymptotically correct (in the large $n$ and
$k$ limit), we can improve the approximation by including more terms
using Gauss-Hermite quadrature. Re-writing the integral again, and
making another change of variables: 
\begin{equation}
	I_{n,k}=\sqrt{\frac{1-\rho}{2\pi\rho|f_{k}''(y_{0})|}}e^{f_k (y_{0})}\int_{-\infty}^{\infty}e^{-\frac{x^{2}}{2} + R_k\left(\frac{x}{\sqrt{|f_{k}''(y_{0})|}}+y_{0}\right)}dx
\end{equation}
where $R_k$ is the remaining terms of $f_k$ after expansion: 
\begin{equation}
	R_k(y)=f_k(y)-f_k(y_{0})+\frac{|f_{k}''(y_{0})|}{2}(y-y_{0})^{2}.
\end{equation}
Now we can approximate the integral using Gauss-Hermite quadrature:
\begin{equation}
	I_{n,k}(N)=\sqrt{\frac{1-\rho}{2\pi\rho|f_{k}''(y_{0})|}}e^{f_{k}(y_{0})}\left[\sum_{i=1}^{N}w_{i}e^{R_{k}\left(\frac{x_{i}}{\sqrt{|f_{k}''(y_{0})|}}+y_{0}\right)}\right],\label{eq:InkN-1}
\end{equation}
where $x_{i}$ are the points for which $H_{N}\left(x_{i}\right)=0$
and the weights $w_{i}$ are 
\begin{align}
w_{i}=\frac{N!\sqrt{2\pi}}{N^{2}\left[H_{N-1}\left(x_{i}\right)\right]^{2}}.\\
\nonumber
\end{align}
Note that $I_{n,k}(1)$ is Laplace's approximation, i.e. Laplace's
approximation is a first order Gauss-Hermite quadrature at the maximum
of $f_k$, while $I_{n,k}(N)$ approximates the remainder terms with
increasingly high order polynomials and so we expect $I_{n,k}(N)\rightarrow I_{n,k}$
as $N$ increases. 

\section{Approximation of inverse cumulative distribution function}

\label{sec:approximation_of_inverse}The normal distribution's inverse
cumulative distribution function, $\Phi^{-1}\left(x\right)$, can be approximated \cite{abramowitz_handbook_2013} for $0<x\leq0.5$ as
\begin{equation}
\Phi^{-1}\left(x\right)\approx\frac{a_{0}+a_{1}s}{1+b_{1}s+b_{2}s^{2}}-s,\quad\quad s=\sqrt{-2\ln\left(x\right)}\label{eq:Phi_inv}
\end{equation}
with \begin{subequations} 
\begin{align}
a_{0} & =2.30753,\quad\quad b_{1}=0.99229,\label{eq:approximation_parameters}\\
a_{1} & =0.27061,\quad\quad b_{2}=0.04481.
\end{align}
\end{subequations}For $0.5<x\leq1$ we use $\Phi^{-1}(x)=-\Phi^{-1}(1-x)$.

\section{Other properties of real networks}

\begin{figure*}
\centering \includegraphics[width=0.9\linewidth]{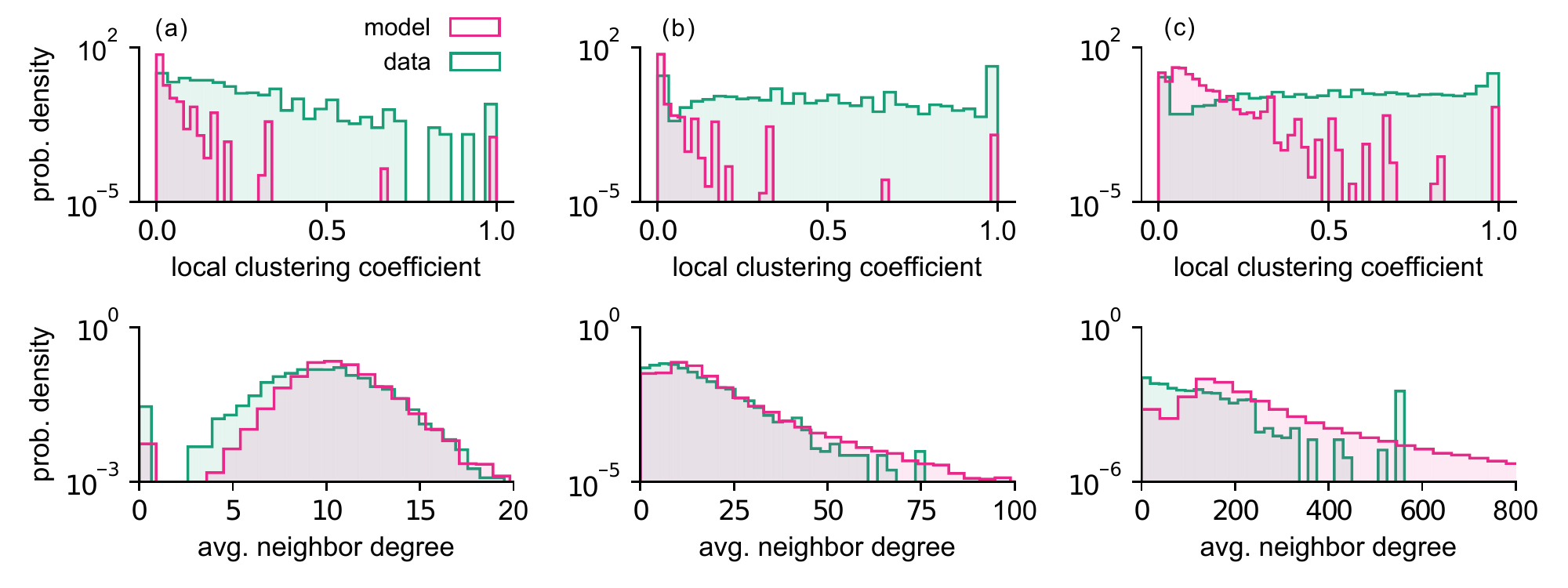}
\caption{Local clustering coefficient and average neighbor degree for the three real-world networks introduced in Sec.~\ref{subsec:degree-distribution} along with 
simulated results from the thresholded normal model. We show (a) a high school friendship
network, (b) a co-authorship network between scientists, and (c) a
protein--protein interaction network.}
\label{fig:real_comparisons-other} 
\end{figure*}

In Fig.~\ref{fig:real_comparisons-other} we compare simulations from the thresholded normal model to the real networks from Sec.~\ref{subsec:degree-distribution}.
Other degree properties, such as the average neighbor-degree seem to be modelled well, while the local clustering coefficient is generally smaller in the simulations than real data, as expected.

\section{Disparity filtering}

The simplistic style of thresholding that we have considered is not the only method to extract a network from relational data.
Indeed, more sophisticated algorithms have been developed \cite{serrano2009extracting, radicchi2011information, dianati2016unwinding}.
By limiting our analysis to a simple thresholding procedure, rather than a more sophisticated algorithm, we have been able to derive several of the basic properties.
In this appendix we repeat some of the analysis for the more complex ``disparity filter'' algorithm \cite{serrano2009extracting}.

The disparity filter assumes that the relational data are positive and vary by orders of magnitude.
In contrast, we have considered data that are distributed according to a standard normal distribution---$X_{ij}$ is negative half of the time and will virtually never be larger in magnitude than $10$.
To match the assumptions of the disparity filter, we apply the algorithm to $e^{X_{ij}}$.

Since $X_{ij}$ is normally distributed, $e^{X_{ij}}$ is log-normally distributed.  And, since the exponential function is monotonic, a simple thresholding procedure at the value of $e^{t}$ is equivalent to our previous analyses.
However, since $e^{X_{ij}}$ is non-negative and will vary by orders of magnitude, these data also match the assumptions of the disparity filter.

The disparity filter algorithm proceeds by assigning a local significance score $\alpha_{ij} \neq \alpha_{ji}$ to each potential edge.  In the normal model this score will be
\begin{equation}
	\alpha_{ij} = \left( \frac{\sum_{k \neq j} e^{X_{ik}}}{\sum_{k} e^{X_{ik}}} \right)^{n-2}.
\end{equation}
An edge is considered present between nodes $i$ and $j$ if either $\alpha_{ij}$ or $\alpha_{ji}$ exceeds a predetermined significance threshold, $\alpha$.

Making use of the decomposition from Eq.~\eqref{eq:WYZ}, we can write
\begin{equation}
	\alpha_{ij} = \left( \frac{\sum_{k \neq j} e^{\sqrt{1 - 2\rho} Y_{ik} + \sqrt{\rho} Z_k  }}{ \sum_{k} e^{\sqrt{1 - 2\rho} Y_{ik} + \sqrt{\rho} Z_k  }} \right)^{n-2}.
\end{equation}
Further, we can make a mean-field style approximation and replace $\sum_{k \neq j} e^{\sqrt{1 - 2\rho} Y_{ik} + \sqrt{\rho} Z_k  }$ by its expected value to arrive at
\begin{equation}
	\alpha_{ij} = \left( \frac{(n-2) e^{\frac{1-\rho}{2}}}{ (n-2) e^{\frac{1-\rho}{2}} + e^{\sqrt{1 - 2\rho} Y_{ij} + \sqrt{\rho} Z_j  }} \right)^{n-2}.
\end{equation}
Defining the constant
\begin{align}
	c &=  \frac{2\log{\left( (n-2)(\alpha^{-\frac{1}{n-2}} - 1) \right) } + (1-\rho) }{2\sqrt{\rho}}
\end{align}
a quick calculation establishes that either $\alpha_{ij}$ or $\alpha_{ji}$ is larger than $\alpha$ with probability
\begin{equation}
	E[A_{ij}] = \int_{-\infty}^{+\infty} \left( 1 - \Phi\left( -x \sqrt{1/\rho-2} + c \right)^2 \right) \phi(x) dx
	\label{eq:disparity_density}
\end{equation}
where $\phi$ and $\Phi$ are again the density function and cumulative distribution function for the standard normal distribution.
This integral can easily be computed using Gauss-Hermite quadrature.
Its derivatives with respect to $\alpha$ are also simple to compute, and so root finding algorithms such as Newton's method can find the correct choice of $\alpha$ for a desired final density.

\begin{figure}[b]
	\centering
\includegraphics[width=3.1in]{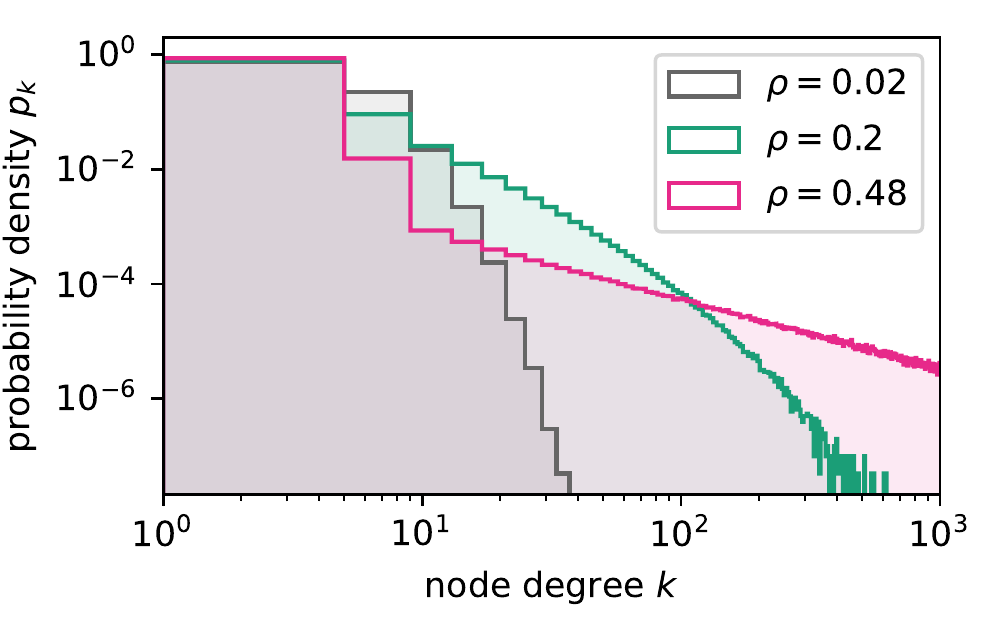}
	\caption{\label{fig:disparity_degrees} Simulations for the disparity filter \cite{serrano2009extracting} applied to $e^{X_{ij}}$. Networks with $2,000$ nodes were created by first sampling matrices $\bm{X}$ from a normal distribution and then applying the disparity filter to $e^{X_{ij}}$.  Three different values for $\rho$ were used and $\alpha$ was set using Eq.~\eqref{eq:disparity_density} so that the final network would have a mean degree of $4$.}
\end{figure}

While the above derivation for the network density in this model is reasonably straight-forward, other properties such as the degree distribution are more challenging.
Instead of re-deriving the results of this paper for a sophisticated filtering algorithm we present the results of simulations in Fig.~\ref{fig:disparity_degrees}.
Increasing the correlation $\rho$ similarly increases the variance on the node degree $k$.
This demonstrates that the qualitative properties derived for the naive thresholding still apply for at least some more sophisticated thresholding methods.

\end{document}